\documentclass[12pt,preprint]{aastex}

\slugcomment{To appear in The Astrophysical Journal (Letters)}

\shorttitle{A Compact, Possibly Isolated Protoplanetary Disk}
\shortauthors{Rodr\'\i guez et al.}

\begin{document}


\title{IRAS~16293-2442B: A Compact, Possibly Isolated Protoplanetary Disk in a Class 0 Object}


\author{Luis F. Rodr\'\i guez, Laurent Loinard, and Paola D'Alessio}
\affil{Centro de Radioastronom\'\i a y Astrof\'\i sica, UNAM,
Apdo. Postal 3-72, Morelia, Michoac\'an, 58089 M\'exico}
\email{l.rodriguez, l.loinard, p.dalessio@astrosmo.unam.mx}

\and

\author{David J. Wilner and Paul T. P. Ho}
\affil{Harvard-Smithsonian Center for Astrophysics, 
60 Garden Street, Cambridge, MA 02138, USA}
\email{dwilner, ho@cfa.harvard.edu}



\begin{abstract}

Theoretical arguments suggest that protoplanetary disks around young stars
should start small and grow with the addition of high angular momentum
material to reach the radii of several hundred AUs that characterize the disks
around optically visible T Tauri stars. Examples of much more compact disks,
with radii much less than 100 AU, have been found around some very young stars,
but in all cases tidal truncation from a near binary companion provides a ready
explanation for the small disk size. We report here an example of a compact,
possibly isolated disk around the class 0 object IRAS16293-2422B, which is
thought to be among the youngest protostars known.  This disk has a Gaussian
half power radius of only $\sim$8~AU, and a detailed, self-consistent, accretion
disk model indicates an outer radius of only 26 AU. This discovery supports
the notion that protoplanetary disks start small and grow with time,
although other explanations for the compact size cannot be ruled out,
including gravitational instability in its outer parts and tidal
truncation from the close approach of a now distant stellar companion.

\end{abstract}





\section{Introduction}

During the last decade there has been an accumulation of images at
radio, infrared, and optical wavelengths
(e. g. Wilner et al. 2000; Padgett et al. 1999; 
Burrows et al. 1996) that substantiate the old 
expectation that young stars form surrounded by circumstellar disks
of gas and dust from which planets may condense in the future.
There has also been significant advance in 
determining the duration of the disk phase, with estimates in
the range of 2 to 10 million years
(Haisch, Lada, \& Lada 2001; Armitage, Clarke, \& Palla 2003),
as well as in studying examples of old disks (e. g. Calvet et al.
2002; Aigen, Lunine, \& Bendo 2003).
However, little is known about the early stages of disk formation,
largely because the disks develop deep within protostellar envelopes
that are opaque except at long wavelengths. Fortunately, the development of the
7 mm system at the Very Large
Array now enables high angular resolution
observations that can penetrate through the
envelopes of Class 0 sources into the inner region where disks form.

Until now, most of the disks with measured dimensions
are associated with Class I or Class II sources
and have radii of order 100 AU or larger
(e. g. Padgett et al. 1999; Kitamura et al. 2002). There 
are theoretical clues that suggest that
the Class 0 sources, the youngest 
protostars known, could possess smaller disks.
However, since these are short-lived
and relatively rare objects, solid observational evidence is
difficult to find and still missing.

The centrifugal radius, the radius at which centrifugal support becomes
important in infalling gas, sets a characteristic
disk size. One expects that the size 
of a disk should be
initially small and increase with time, as gas with progressively higher angular
momentum falls into the disk (Terebey, Shu, \& Cassen 1984;
Ruden \& Lin 1986; Lin \& Pringle 1990).

Obviously, direct imaging of young stars in various evolutionary stages
would be the best way to investigate if disks grow in size during the
early stages of star formation. A handful of compact (i. e. with radii 
much smaller than 100 AU) disks have been imaged toward
the young stars L1551~IRS5 (Rodr\'\i guez et al. 1998; radii of
$\sim$10 AU),   
L1527 (Loinard et al. 2002; radius of $\sim$ 20 AU), and SVS~13 (Anglada et al. 2004;
radius of $\leq$30 AU). 
However, in these cases a nearby companion has always been found that
could be making the disks small by tidal truncation.

In this paper we report 7 mm observations of
IRAS~16293-2422B, a class 0 object located in the Ophiuchus molecular complex
at a distance of 120 pc (Knude \& Hog 1998). Class 0 objects
are extremely young stars, with an estimated age of less
than $\simeq 10^4$ yr (Andr\'e, Ward-Thompson, \& Barsony 1993). Our observations
reveal the first case of
an isolated, compact protoplanetary disk around 
a class 0 star, a result that supports the hypothesis that
disks start small and grow with evolution.
 
\section{Observations}


The 7 mm observations were carried out in 2003 June 22
using the Very Large Array (VLA) of the National
Radio Astronomy Observatory (NRAO)\footnote{NRAO is a facility of the
National Science Foundation operated under cooperative agreement by
Associated Universities, Inc.}\ . The array was then in the A configuration
and we had excellent phase stability during the run.
The longest baselines of the array reached $\sim$36 km, while the
shortest ones were $\sim$0.7 km
These observations have an angular resolution of $\sim 0\rlap.{''}07$,
unmatched at present in the millimeter regime, and provide detailed imaging
of circumstellar structures.
 
For all observations we used the fast-switching
technique that consists of rapidly alternating observations of the
source and the phase calibrator with cycle time of 2 minutes.
An effective bandwidth of 100~MHz with two circular
polarizations was employed. The absolute amplitude calibrator was
1328+307 (adopted flux density of 1.45 Jy), used in
conjunction with a source
model provided by NRAO. The phase calibrator was
1622$-$253, with a bootstrapped flux density of 1.79$\pm$0.02 Jy.
The data
were edited and calibrated using the software package Astronomical Image
Processing System (AIPS) of NRAO. Cleaned maps were obtained using the
task IMAGR of AIPS and natural weighting.
An extended source was detected in association with the source
IRAS~16293-2422B (see Figure 1).




\section{Discussion}

\subsection{Dust Spectrum}

The structure detected in association with IRAS~16293-2422B has a
total flux density of 31$\pm$1 mJy, as determined from the cumulative
flux in annulli centered on the source. The error given for the total
flux density is a formal error; the total error is dominated by
systematics at a $\sim$5\% level.

To construct the spectral energy distribution of IRAS16293B, we
combined our new 7 mm observation with data taken from the literature.
For the 2.7 mm (107.75 GHz) point, we used the value obtained by
Looney, Mundy \& Welch (2000) when they restore their data to be most
sensitive to compact structures, since this is the situation most
similar to ours. It should be pointed out, however, that the 2.7 mm
flux density does not change much when all angular scales are taken
into account (it merely increases from 0.42 to 0.55 Jy). This suggests
that as much as 75\% of the millimeter emission in IRAS16293B does
emanate from the compact disk seen here at 7 mm.  The 0.8 mm (354 GHz)
data point was obtained by Kuan et al.\ (2004) with an angular
resolution of $\sim 2{''}$. The centimeter flux densities were
obtained from Wootten (1989),
Loinard (2002), Estalella et al. (1991), and Mundy et al. (1992).

The spectrum shown in Figure 2 can be fitted with a single power law,
$S_\nu \propto \nu^{2.6}$, for frequencies in the range of 8.45 to
107.75 GHz, indicating that dust emission is dominant even at the
relatively low frequency of 8.45 GHz. It is only below 8.45 GHz that a
significant free-free excess is evident. Estalella et al. (1991) had
reached similar conclusions. We then conclude that the 7 mm data point
falls in the wavelength regime dominated by dust.

\subsection{Disk Structure}

A Gaussian least-squares fit to the source gives 
a deconvolved HPFW of $\sim 0\rlap.{''}14$,
with no clear indications of significant ellipticity.
At a distance of 120 pc, this corresponds to a 
characteristic half power radius of about 8 AU.
This value is much smaller than the radii of several hundred AU
that have been reported in other more evolved sources (e. g. 
Yokogawa et al. 2001; Jayawardhana et al. 2002;
Wolf, Padgett, \& Stapelfeldt 2003).
The deconvolved HPFW of the 7 mm emission of two disk sources at comparable distances,
HL~Tau and TW Hya,
obtained with the same techniques described here
(Wilner, Ho, \& Rodr\'\i guez 1996; Wilner et al. 2000),
correspond to characteristic radii of $\sim$50 AU, significantly
larger than the value of $\sim$8 AU determined for IRAS~16203-2422B. 
The Gaussian least-squares fit accounts well for the emission within
$\leq 0\rlap.{''}2$ from the center of the disk.
However, there is faint, extended emission associated with the
source. Of the total flux density of 31$\pm$1 mJy, the Gaussian fit accounts
only for 25$\pm$1 mJy, suggesting that $\sim$6 mJy,
about 20\% of the total flux density,
comes from a structure at radii larger that $0\rlap.{''}2$. If this extended
emission had circular symmetry, as the compact, central part of the source,
we would have to reconsider the extent of the disk. However, our analysis
clearly indicates that the extended emission is not circularly symmetric
and must have a different origin than the disk. The lack of
symmetry in the extended component is marginally evident in Figure 1,
where we see faint emission beyond $0\rlap.{''}2$ from the center of the disk,
to the NE and SW. A more conclusive way of studying the morphology of this
faint extended emission is by determining the average flux density beam$^{-1}$
in ``slices'' taken at different position angles
in a ring going from $0\rlap.{''}2$ to $0\rlap.{''}6$. The inner ring
radius was selected because the compact, bright disk seems to become
undetectable at this radius in the 7 mm image (see Figure 1), and the
outer ring radius was selected because the faint,
extended structure seems to stop contributing
significantly to the integrated flux density (see discussion
and Figure 4 below). In Figure 3
we show these average flux densities as a function of
position angle, where it is evident that there
is significant emission at position angles of $\sim 70^\circ \pm 45^\circ$
and $\sim 250^\circ \pm 45^\circ$, and little or no emission outside
of these position angle ranges. At present the nature of this extended
emission is unclear. It could be related to the complex outflow
phenomena associated with this source, although the bipolar outflow associated
with IRAS~16293-2422B is at a position angle of $\sim 110^\circ$  
(Hirano et al. 2001).

\subsection{Isolated Nature of the Source}

We then conclude
that the IRAS~16203-2422B disk is much 
smaller than the cases previously mentioned and
more comparable in size with the tidally truncated disks found in
young close binary systems (L1551~IRS5: Rodr\'\i guez et al. 1998;   
L1527: Loinard et al. 2002; SVS~13: Anglada et al. 2004).
However, in IRAS~16203-2422B the nearest stellar object known is IRAS~16293-2422A
(possibly a close binary system on its own, see Loinard 2002), that at
a projected distance of 600 AU, appears incapable at present of explaining the small
size of the disk in IRAS~16293-2422B by tidal truncation.
For tidal truncation to be effective, the separation between the stars
should be only a few times the radius of the disk (Artymowicz \& Lubow 1994).
However, IRAS~16203-2422A
and IRAS~16203-2422B may form a system of high eccentricity that comes very close
at periastron. The eccentricity
of an elliptical orbit is
$e$ = ($d_{a}$ - $d_{p}$)/($d_{a}$ + $d_{p}$),
where $d_{a}$ is the separation between stars at apastron and $d_{p}$ is the
separation between stars at periastron.
Assuming that the present observed projected separation between
components A and B, $d_{obs} \simeq$ 600 AU, is a lower limit for
$d_{a}$ and that the required
separation for tidal truncation to be effective,
$d_{tidal} \simeq$ 100 AU, is of order of $d_{p}$,
we find that the required eccentricity must be in the range
$0.7 \leq e < 1.0$, making this explanation somewhat unlikely.
It is interesting to point out that the image of IRAS~16203-2422B shown
in Figure 1 has marginal evidence of a spiral-like structure, a
phenomenon that can result from tidal interaction (Grady et al. 2001;
Clampin et al. 2003; Artymowicz \& Lubow 1994).
On the other hand, IRAS~16293-2422B is firmly established as a
class 0 object (Andr\'e, Ward-Thompson, \& Barsony 1993) with an age of less than
$10^4$ years and we tentatively attribute the compactness
of its disk to the youth of the source.

\section{Disk model}

The Gaussian half power size is an underestimate of the outer radius of the
disk because of the central
rise of disk surface density and temperature (Mundy et al. 1996), and 
with the purpose of estimating 
more accurately the basic parameters of the disk, our 
observations are compared to the model of 
a steady irradiated $\alpha$-disk, of the kind usually 
used to describe disks around Classical T Tauri Stars 
and Herbig Ae/Be stars.
The details of how the model is calculated are given elsewhere 
(D'Alessio et al. 1998, 1999, 2001, 2004). In summary, 
the disk is  assumed to have a constant mass accretion rate, $\dot{M}$,  
and a viscosity coefficient given by $\nu_{turb}= \alpha c_s H$, where 
$\alpha$ is a free parameter, taken to be constant 
through the disk, $c_s$ is the local sound speed and 
$H$ is the local gas scale height. The disk is in Keplerian rotation 
and its self-gravity is neglected compared to the stellar gravity.
In an $\alpha$-disk,  the $\dot{M}$ and the parameter $\alpha$ 
are  related to the disk mass surface density through the 
conservation of angular momentum flux, given by  

\begin{equation}
\Sigma \approx {\dot{M} \Omega_k \over 3 \pi \alpha c_s^2} 
\biggl [ 1 - \biggl({R_* \over R} \biggr)^{1/2} \biggr].
\end{equation}

We assume the central star 
has a mass $M_*=0.8 \ M_\odot$ (Ceccarelli et al. 2000), 
and we adopt for it an effective temperature 
 $T_* \sim $ 4000 K, a stellar luminosity $L_* \sim 10 \ L_\odot$ 
and a radius $R_* \sim 6.7 \ R_\odot$, roughly consistent with 
pre-main sequence tracks in the HR diagram (Siess, Dufour \& Forestini 2000).
Adopting a system bolometric luminosity of
$L_{bol} \sim 16 \ L_\odot$, the accretion luminosity 
is $L_{acc} =G M_* \dot{M}/R_* \sim  6 \ L_\odot$, 
implying a mass accretion rate $\dot{M} \sim 1.7 \times 10^{-6} \ 
M_\odot$year$^{-1}$.
All these values are very uncertain, thus, the disk model we present 
here has only an illustrative purpose, mainly to show that
the disk interpretation is consistent with the data.

The disk is being heated by local viscous dissipation and by 
external irradiation. 
Since the stellar and accretion luminosities are comparable, 
we assume that the disk is being irradiated by both, the central star and
the accretion shocks at the stellar surface, where most of the 
accretion luminosity is released.
For simplicity, the shocks are  assumed to emit as a blackbody 
with $T_{shock} \sim 8000$ K (Calvet \& Gullbring 1998;
Gullbring et al. 2000) and to cover a few percent  of the stellar surface 
to account for the assumed accretion luminosity.

The main opacity source of the disk is dust, and the disk structure and 
emergent intensity depends on the grain's composition and size distribution.
We assume that the disk has interstellar-like dust grains in 
its upper layers and a population of millimeter-sized grains in its
midplane.
The small grains at the disk atmosphere absorb efficiently radiation
from the star and the accretion shocks, characterized by shorter wavelengths
than the disk's own radiation. On the other hand, the millimeter-sized grains
at the midplane have a larger emissivity at $\lambda \sim $ mm than 
interstellar medium grains. 
This model with small and big dust grains  
has a larger emergent intensity at 7 mm than a model with only 
millimeter-sized grains everywhere, since such big grains are less 
effective for absorbing stellar radiation
and heating the disk (D'Alessio et al. 2004). 
The steady disk models that fit better the observations shown in Figure
4 have a maximum radius $R_d = $ 26 AU and a mass in 
the range $M_d=0.3- 0.4 \ M_\odot$.

For larger distances, the contribution of the extended component
described in \S 3 becomes important, accounting for $\sim$ 6 mJy
at $0\rlap.{''}6$ = 72 AU. Figure 4 shows that this extended component
has a cumulative flux density that does not grow as fast with
distance as what an extrapolation of the steady disk model predicts 
(roughly $F \propto R$).
One possibility is that this extended component is the outer disk
where the mass surface density decreases exponentially with radius, according
to evolutive models that take into account the disk expansion to
conserve the total angular momentum (Hartmann et al. 1998).
Another possibility is that our steady model breaks down in the
outer regions, because the disk self-gravity becomes important.
One way to evaluate this is by calculating the Toomre parameter 
corresponding to the disk model which fits the observations,
$Q= c_s \ \Omega_k/ 2 \pi G \Sigma$, where we take the sound speed
$c_s$ evaluated at the disk midplane. For $R \sim $ 26 AU,
$Q \sim 0.4-0.3$, implying that the outer disk is gravitationally unstable.
This instability might be the origin of the non-axisymmetric structures
we find in the extended region (e.g. Gammie 2001, Rice et al. 2003).
Models in which self-gravity is not neglected and
where the outer exponential-$\Sigma$ region is included, should be used
to further explore these possibilities.

\section{Conclusions}

In conclusion, the disk imaged around IRAS~16293-2422B
appears to be much smaller than the disks detected in more evolved
young stars. The lack of a nearby detectable companion suggests that
this compactness is a result of the youth of the object and not of
tidal truncation. However, we cannot rule out other possible
mechanisms that may explain the compactness of this disk.
For instance, the outer parts of the disk may be gravitationally unstable
and may have collapsed to clumpy structures more difficult
to detect. Also, the source IRAS~16293-2422A appears now to
be too far from IRAS~16293-2422B to produce tidal truncation
but it may approach significantly during periastron. More
extensive, high angular resolution imaging of disks 
around young stars in various evolutionary stages are needed to
critically test the possibilities hinted by our observations.



\acknowledgments

LL, LFR, and PD are grateful to 
CONACyT, M\'exico and DGAPA, UNAM for their support.





\clearpage



\begin{figure}
\epsscale{1.0}
\plotone{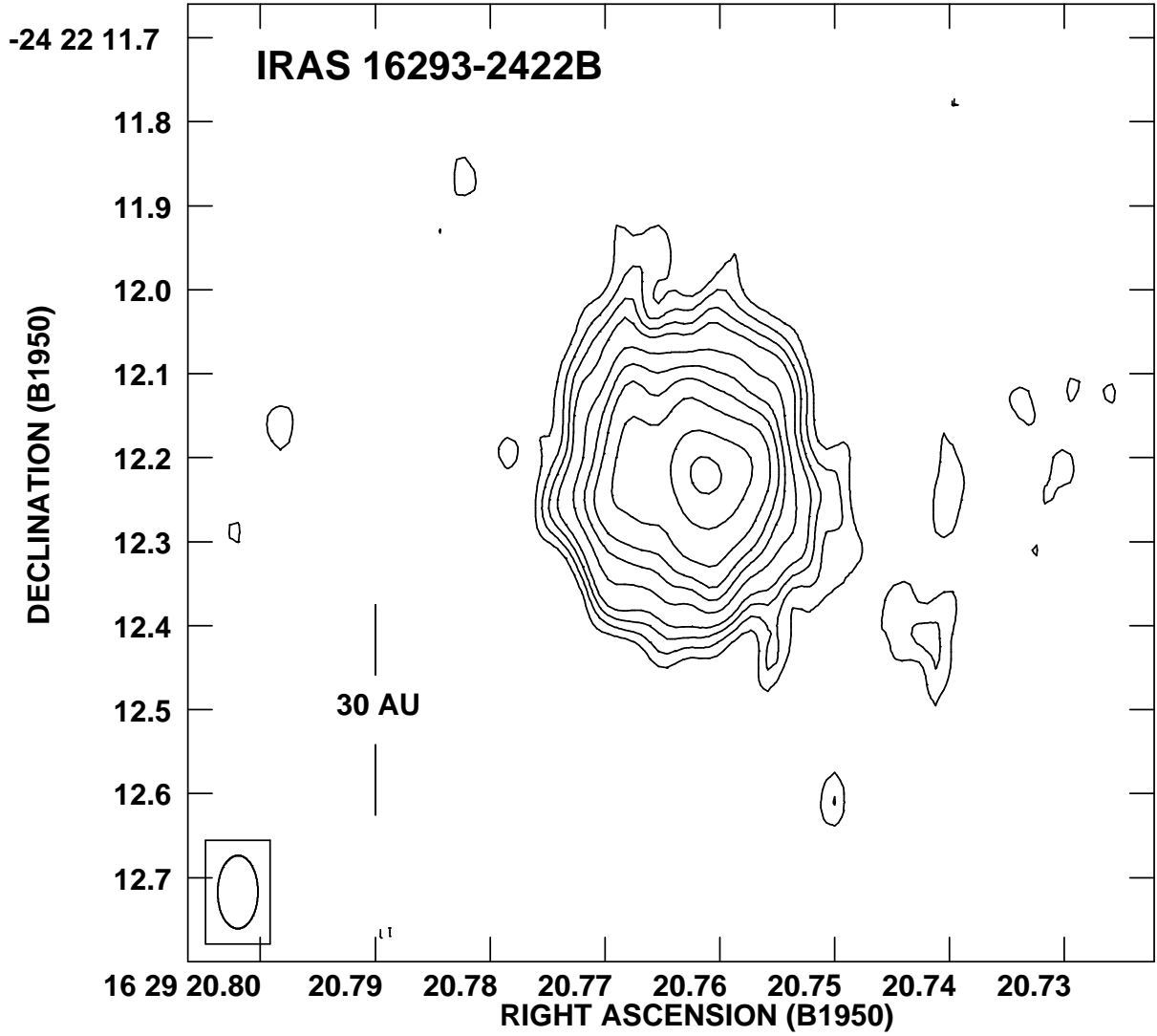}
\caption{VLA image at 7 mm of IRAS~16293-2422B.
Contours are -3, 3, 4, 5, 6, 8, 10 ,12, 15, 20, and 25 times
0.1 mJy beam$^{-1}$, the rms of the image.
The half power contour of the synthesized beam
($0\rlap.{''}09 \times 0\rlap.{''}05$; PA = $0^\circ$) is shown in the bottom left corner.
\label{fig1}}
\end{figure}

\clearpage


\begin{figure}
\centering
\includegraphics[angle=-90,scale=1.0]{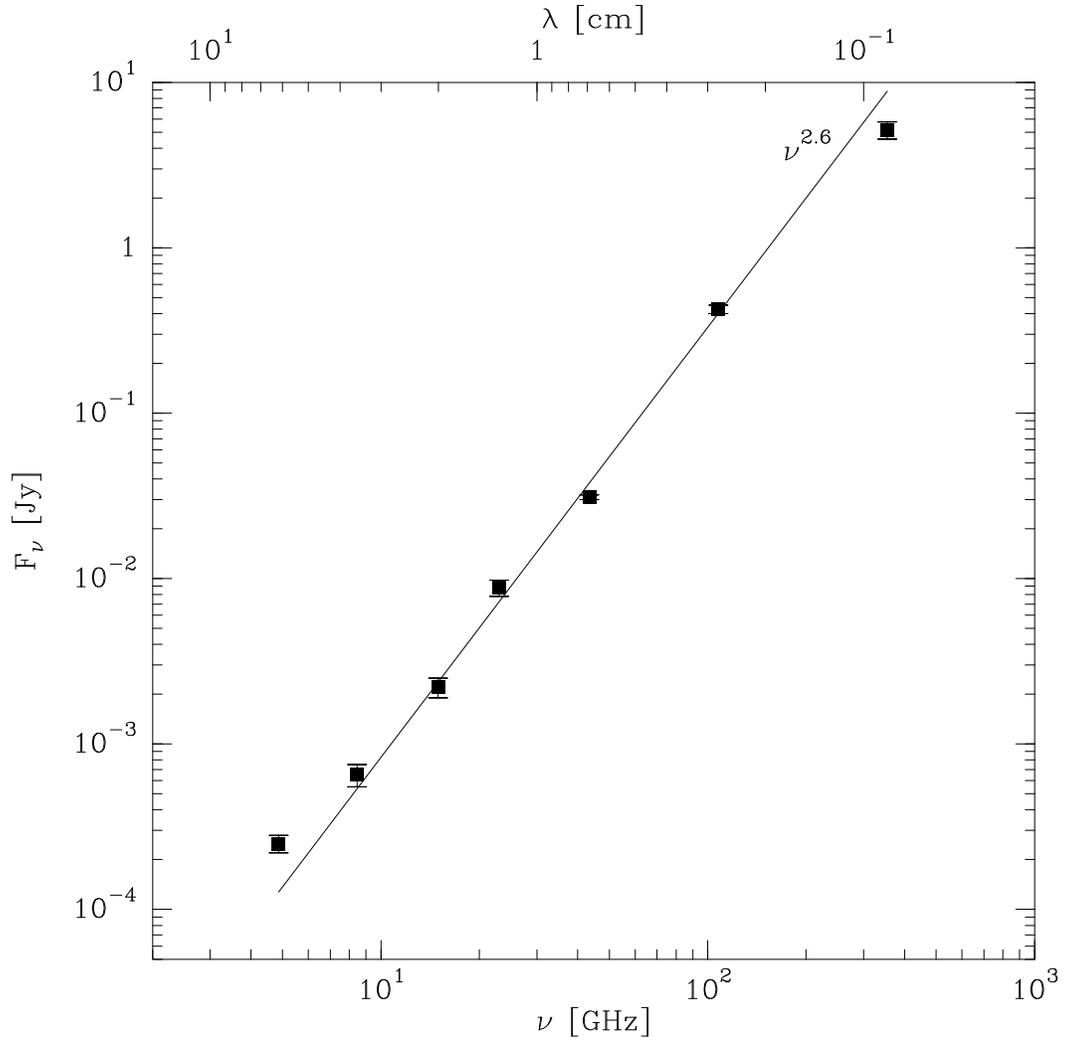}
\caption{Centimeter and millimeter spectrum of IRAS~16293-2422B.\label{fig2}}
\end{figure}


\begin{figure}
\epsscale{1.0}
\plotone{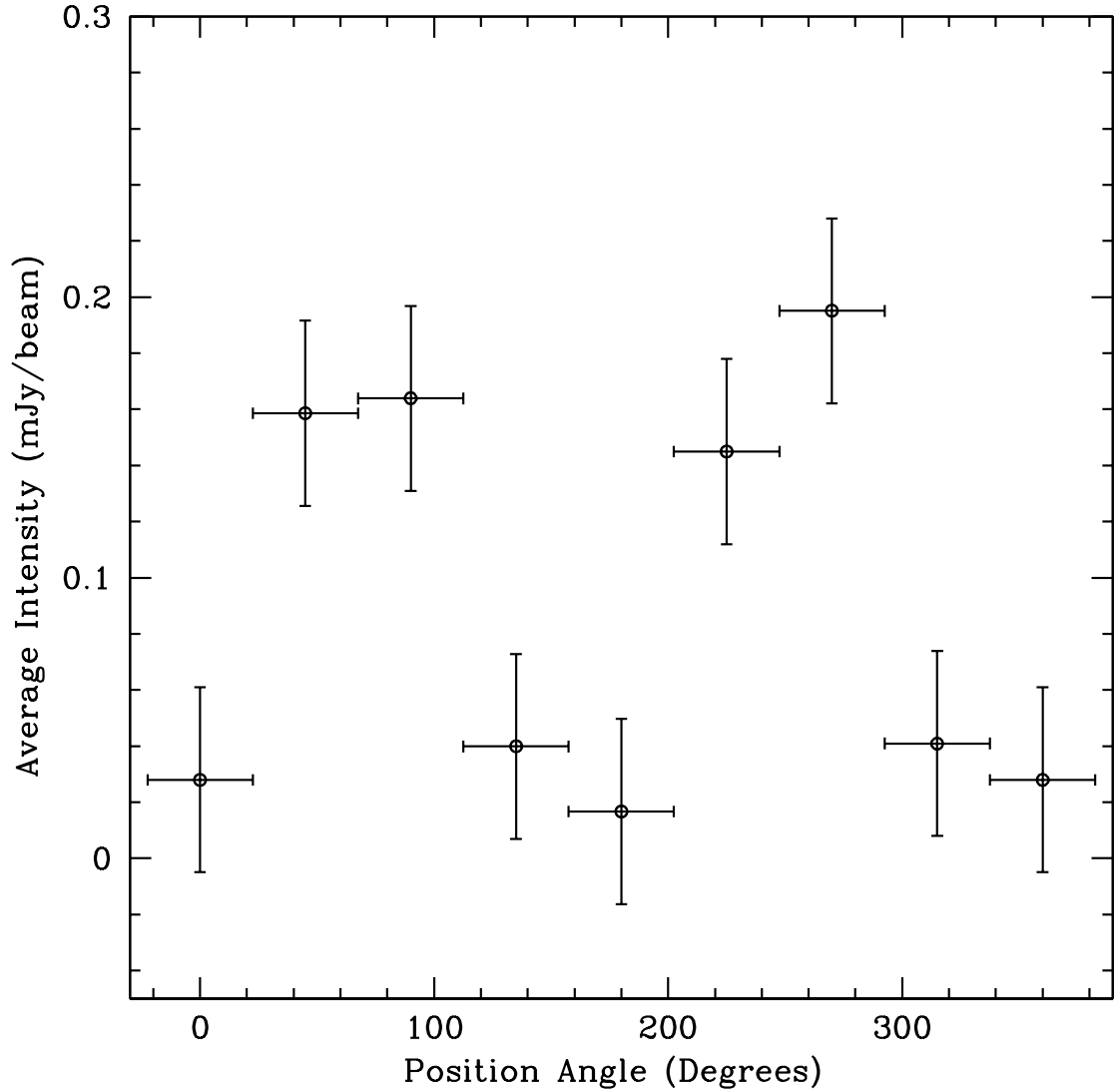}
\caption{Average intensity (mJy beam$^{-1}$) as a function
of position angle (degrees) in a ring with radius from $0\rlap.{''}2$ to
$0\rlap.{''}6$ for the 7 mm source associated
with IRAS~16293-2422B. Note the presence of emission at $\sim 70^\circ \pm 45^\circ$ and
$\sim 250^\circ \pm 45^\circ$.
\label{fig3}} 
\end{figure}

\begin{figure}
\epsscale{1.0}
\plotone{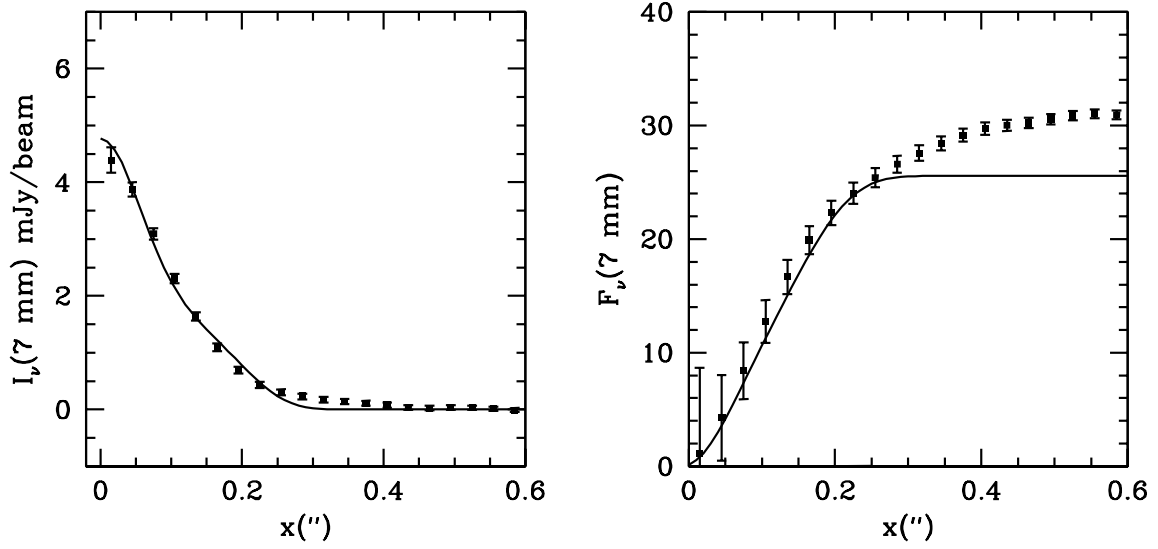}
\caption{The left panel shows  the
radial profile of the azimuthally averaged emergent intensity at 7 mm,
calculated adopting a $0\rlap.{''}1$ circular beam.
The right panel shows the cumulative flux at 7 mm as a function 
of the radial distance to the center of the disk.
\label{fig4}}
\end{figure}






\clearpage

\end{document}